\documentclass[12pt]{iopart}
\usepackage{graphicx}
\usepackage{subfigure}
\usepackage{caption}
\usepackage{lineno}
\usepackage{color}
\usepackage[colorlinks,
linkcolor=blue,
anchorcolor=red,
citecolor=blue
]{hyperref}

\newcommand{\ba}[0]{\begin{eqnarray}}
\newcommand{\na}[0]{\end{eqnarray}}

\begin{document}
	
\title[]{Impact of edge-localized resonant magnetic perturbation on L-H transition in the KSTAR tokamak}
	
\author{Y. Zhang$^1$, M. Jiang$^{1,\ast}$, K. X. Han$^1$, W. H. Ko$^{2}$, S. M. Yang$^{3}$, J. D. Hu$^4$, Z. B. Guo$^{5}$, M. J. Choi$^{2}$, J. W. Juhn$^{2}$, J. Lee$^{2}$,  J. H. Lee$^{2}$, Y. H. Lee$^{2}$, Y. F. Liu$^1$, X. B. Li$^4$, Y. R. Zhu$^{1}$, G. Q. Dong$^{1}$ and G. Z. Hao$^{1}$}

\address{$^1$ Southwestern Institute of Physics, P. O. Box 432, Chengdu 610041, People’s Republic of China}
\address{$^2$ Korea Institute of Fusion Energy, Daejeon 34133, Republic of Korea}
\address{$^3$ Princeton Plasma Physics Laboratory, Princeton, NJ 08543, USA}
\address{$^4$ State Key Laboratory of Nuclear Physics and Technology, Fusion Simulation Center, School of Physics, Peking University, Beijing 100871, People’s Republic of China}
\address{$^5$ Institute of Modern Physics, Fudan University, Shanghai 200433, People’s Republic of China}

\ead{jiangm@swip.ac.cn}

\vspace{10pt}
\begin{indented}
\item[]\date{\today}
\end{indented}
	
\begin{abstract}
	The impact of edge-localized resonant magnetic perturbation (ERMP) on the low-to-high (L-H) confinement mode transition is investigated through dedicated experiments on the KSTAR tokamak with tungsten divertor and numerical simulations with a one-dimensional (1D) reduced transport model. Experimentally, under matched conditions, the ERMP inhibits the L-H transition, whereas the reference discharge without ERMP enters H-mode. The ERMP enhances edge electron temperature fluctuations and the turbulence decorrelation rate, and flattens the toroidal rotation, reducing the edge $E\times B$ velocity shear below the threshold required for turbulence suppression. An extended 1D predator-prey model, which incorporates toroidal flow evolution and realistic radial profile of the magnetic perturbation due to plasma response field, reproduces the experimental trends of enhanced turbulence, reduced $E\times B$ flow shear, and inhibited L-H transition. These findings contribute to a better understanding of how ERMP affects L-H transition dynamics in tungsten divertor environments and offer a valuable reference for the application of ERMP in current and future fusion devices.
\end{abstract}

%
%
%
%
%

\section{Introduction}

High confinement mode (H-mode) is the reference operating scenario for ITER and future tokamak reactors, characterized by the formation of an edge transport barrier (ETB) that substantially improves global energy confinement \cite{Wagner_1982, Wagner_1984}. However, H-mode operation is intrinsically accompanied by edge localized modes (ELMs), which release transient bursts of heat and particle flux onto plasma-facing components and challenge the heat load tolerance of divertor targets \cite{Leonard_2014}. Resonant magnetic perturbations (RMPs) have emerged as an effective external actuator for mitigating or suppressing ELMs, and have been applied on DIII-D, KSTAR, ASDEX Upgrade, EAST, and HL-2A/HL-3 \cite{Evans_2004, Jeon_2012, Suttrop_2017, Sun_2016, Sun_2021,Wang_2024}. As the first large ELM may occur immediately after the L-H transition, RMPs are typically required to be activated during the L-mode phase preceding the transition. This operational requirement creates an inherent coupling between RMP application and the L-H transition dynamics that must be understood for reliable H-mode access.

It has been widely observed across multiple devices that RMPs raise the L-H transition power threshold \(P_{\mathrm{LH}}\), effectively inhibiting H-mode access if the heating powers are not sufficient \cite{Leonard_1991, Kirk_2011, Kaye_2011, Gohil_2011, Ryter_2012, In_2017, Schmitz_2019, Willensdorfer_2022,Zhang_2025}. Systematic experiments have established that the threshold increase is not universal but emerges when the perturbed magnetic field exceeds a critical value, above which \(P_{\mathrm{LH}}\) rises with the strength of magnetic field perturbation \cite{Schmitz_2019,Willensdorfer_2022,Park_2019}. The physical picture attributes this degradation primarily to a reduction of the edge $E\times B$ flow shear \cite{Yoshida_2025}. It has been demonstrated on DIII-D that \(n=3\) ($n$ is the toroidal mode number) RMP reduces the flow shearing rate and increases the turbulence decorrelation rate in the L-mode edge, disrupting the turbulence suppression mechanism that triggers the L-H transition \cite{Schmitz_2019,Mordijck_2016,Kriete_2020}. In addition, RMP decreases the Reynolds-stress-driven energy transfer from turbulence to flows, challenging the energy depletion picture of the L-H bifurcation \cite{Kriete_2020}. Moreover, experimental results on the ASDEX Upgrade shows that RMP can invert the edge negative radial electric field well and enhance turbulence for resonant perturbations \cite{Conway_2015}. These findings converge on the conclusion that RMP does not merely add a transport channel but actively reorganizes the turbulence-flow coupling that governs the transition.

The physical mechanisms through which RMP alters the edge turbulence-flow system can be grouped into two categories. First, stochastic magnetic fields dephase the turbulence-zonal flow coupling: the random walk of field lines decorrelates turbulent wave packets, reduces the efficiency of Reynolds stress drive, and thereby weakens zonal flow generation \cite{Leconte_2014, Chen_2021}. This dephasing effect is the dominant mechanism in the L-mode phase and explains why RMP raises the L-H threshold before the transition occurs. Second, resonant Maxwell stress from the plasma response to RMP can produce an additional electromagnetic torque on the toroidal flow \cite{Diamond_2013,Guo_2022}. This multi-mechanism picture has motivated the development of reduced transport models that incorporate RMP effects through a dephasing factor in the turbulence-zonal flow equations and through explicit Maxwell stress torque terms in the toroidal momentum equation \cite{Miki_2012, Miki_2013, Li_2024,Zhang_2025,Hu_2026s}.

A critical gap in the current understanding concerns the role of divertor material and its interplay with RMP-driven edge transport. Most of the experimental evidence establishing the RMP-\(P_{\mathrm{LH}}\) correlation was obtained in carbon-wall devices, where the L-H threshold itself is influenced by the recycling and impurity dynamics characteristic of carbon surfaces. Tungsten divertors, now adopted in KSTAR, ASDEX Upgrade, and ITER, present a fundamentally different edge environment with higher sputtering thresholds, lower fuel retention, and different core-edge coupling \cite{Park_2019}. Recent KSTAR experiments in the newly upgraded tungsten lower single-null (LSN) divertor have revealed that the L-H transition power threshold exhibits substantial scatter, with different power ramping scenarios acting as a hidden variable \cite{Kim_2025_APS}. In addition, edge-localized RMPs (ERMPs) have been found to affect the L-H transition in multiple ways: they can delay the transition to H-mode, suppress ELMs, drive a back transition from H-mode to L-mode, introduce hysteresis, and alter the magnetic fluctuation spectrum \cite{ Kim_2025_IAEA}. A dedicated turbulence statistical analysis further shows that ERMPs influence the L-H transition by disrupting coherent modes and promoting bursty activity \cite{Kim_2025}. These observations indicate that ERMPs influence the transition dynamics through a richer set of mechanisms than previously recognized. The KSTAR results also show that the impact of ERMPs is sensitive to plasma density, with the threshold behavior differing between the tungsten LSN and carbon upper single-null (USN) configurations \cite{ Kim_2025_IAEA}. These observations raise the question of whether the fundamental turbulence-flow mechanism established in carbon devices remains quantitatively valid in a tungsten environment, where impurity radiation, neoclassical transport, and edge profile gradients differ systematically.

From a modeling perspective, significant progress has been made in reduced models that capture RMP effects on the L-H transition. A zero-dimensional predator-prey model incorporating stochastic fields demonstrated that RMP raises the transition threshold by quenching zonal flow generation through vorticity flux decoherence \cite{Leconte_2014}. The extension to one-dimensional (1D) models has enabled the treatment of radially resolved profiles and the identification of the power threshold scaling with the amplitude of magnetic field perturbation \cite{Zhang_2025,Hu_2026s}. However, existing 1D studies have generally adopted assumed magnetic perturbation profiles rather than radial distributions derived from the plasma response field, and they have not been fully tested with experimental parameters to see whether they can qualitatively and quantitatively reproduce and explain the relevant experimental observations.

In this paper, we investigate the impact of ERMP on the L-H transition on the KSTAR tokamak with LSN tungsten divertor. We combine dedicated KSTAR experiments with a 1D predator-prey model that incorporates the ERMP-induced magnetic field perturbation through its effects on turbulence-zonal flow dephasing and on the toroidal rotation evolution. Section \ref{Exp} describes the experimental setup and the key observations on KSTAR, including the comparison of discharges with and without ERMP under matched plasma parameters and heating power. Section \ref{Sim} presents the 1D predator-prey simulation framework, the construction of the magnetic field perturbation profile from plasma response calculations, and the comparison between simulation results and experimental measurements of the edge $E\times B$ flow shear. Section \ref{Con} summarizes the conclusions and discusses the potential future works.

\section{Experimental setup and results on KSTAR tokamak\label{Exp}}

\subsection{Experimental setup and diagnostics}

Experiments were performed on the KSTAR tokamak \cite{Nam_2026}, a superconducting device with major radius $R_0 = 1.8 m$, minor radius $a = 0.5 m$, toroidal magnetic field up to $B_T = 3.5 T$, and plasma current up to $I_p = 2 MA$. The heating and current drive system consists of neutral beam injection (NBI) and electron cyclotron heating (ECH), with a total installed power of up to $\sim 14 MW$ NBI and $\sim 2 MW$ ECH \cite{Nam_2026}. For the present study, discharges were conducted in a LSN configuration with the newly installed tungsten mono-block divertor. Typical plasma parameters were plasma current $I_p = 0.6 MA$, toroidal magnetic field $B_T = 1.8 T$, elongation $\kappa = 1.9$, upper triangularity $\delta_u = 0.32 $, lower triangularity $\delta_l = 0.75$, edge safety factor $q_{95} \sim 4.6$, and L-mode line-averaged density $n_e \sim 3.1 \times 10^{19} m^{-3}$.

The electron density and temperature profiles were measured by the Thomson scattering (TS) diagnostic \cite{Kim_2025TS}, with a time resolution of $50 ms$ and spatial resolution of  $0.5 cm$ in the edge and $2 cm$ in the core. The ion temperature and toroidal rotation velocity profiles were obtained from charge exchange recombination spectroscopy (CES) \cite{Ko_2010}, with a time resolution of $10 ms$ and spatial resolution of $0.5 cm$ in the edge and $2 cm - 3 cm$ in the core. The KSTAR CES system measures the charge exchange spectra of the carbon VI impurity line at $529 nm$ to obtain ion temperature and rotation profiles. The two-dimensional electron temperature fluctuations were measured by the electron cyclotron emission imaging (ECEI) diagnostic \cite{Yun_2010}, which provides high spatial resolution of $1 cm -2 cm$ and temporal resolution of $0.5 \mu s - 2 \mu s$. 

Specifically, RMPs were applied using the ITER-like three-row in-vessel control coils (IVCCs) \cite{In_2019}. The IVCC system consists of three rows of coils (top/middle/bottom) that allow flexible control of the 3D field spectrum with toroidal mode numbers $n = 1, 2, 3$. The ERMP is an optimized RMP spectrum that minimizes the core resonant response while maintaining sufficient edge resonant field for ELM control \cite{Yang_2020,Yang_2024}. In this study, the ERMP was applied with $n = 1$ and a coil current of up to $1.5 kAt$, producing a relative vacuum radial magnetic field perturbation at $q_{95}$ surface of $\delta B_r/B_T \sim 5.5 \times 10^{-4}$. The plasma response to the applied field was calculated using the ideal perturbed equilibrium code (IPEC) \cite{Park_2007,Yang_2020} to assess the realistic amplitude of the  perturbed radial magnetic field ($\delta B_r/B_T \sim 2.24 \times 10^{-4}$ at $q_{95}$ surface).

\subsection{Experimental results}

\begin{figure}[htb!]
	\centering
	\includegraphics[width=\linewidth,scale=1.0]{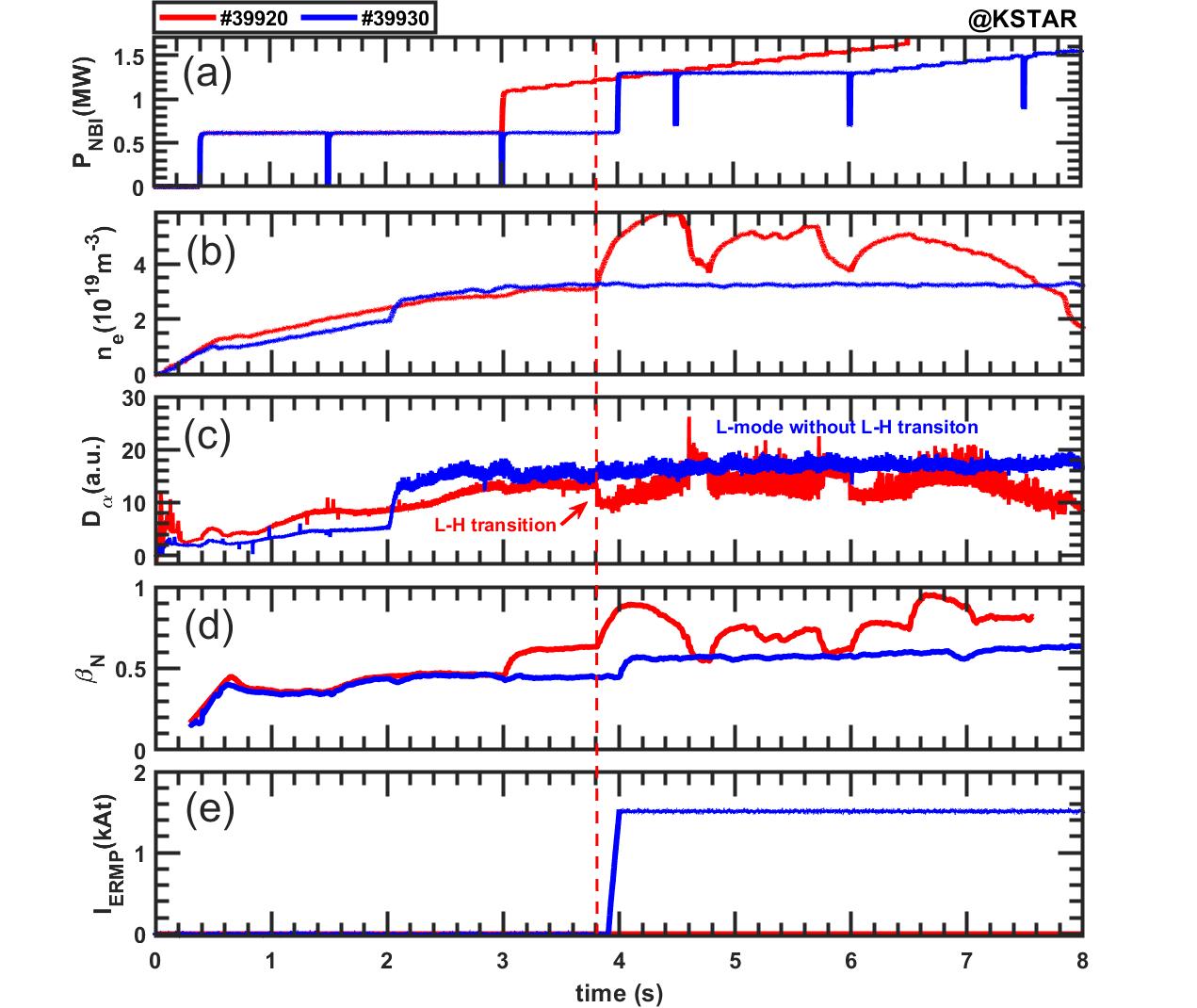}
	\caption{ Time evolutions of the discharge parameters in shot 39920 and 39930, with (a) the NBI heating power $P_{NBI}$, (b) the line-averaged electron density $n_e$, (c) the $D_\alpha$ intensity in the divertor, (d) the normalized beta $\beta_N$, and (e) the ERMP current $I_{ERMP}$. The red vertical dashed line marks the L-H transition time in shot 39920.}
	\label{Diagram_39920_39930_modified} 
\end{figure}

Figure \ref{Diagram_39920_39930_modified} compares the main discharge parameters of two shots, \#39920 and \#39930, which were performed under similar experimental conditions. Shot \#39920 serves as the reference discharge without ERMP. Its NBI heating power increases linearly from $1 MW$ at $t=3 s$ at a rate of $\sim 0.18 MW/s$, while the line-averaged electron density remains approximately $3.1 \times 10^{19} m^{-3}$. When the NBI power reaches $P_{NBI}=1.2 MW$ at $t=3.8 s$, the line-averaged density and normalized beta $\beta_N$ begin to rise rapidly, accompanied by a sudden drop in the $D_\alpha$ signal. These signatures indicate the occurrence of the L-H transition. In contrast, in shot \#39930 the $n=1$ ERMP is applied from $t = 4 s$ with a constant current of $I_{ERMP} = 1.5 kAt$. A constant NBI heating power of $1.25 MW$ is maintained from $t= 4s$ to $t= 6s$, after which it increases linearly at a rate of $\sim 0.15 MW/s$. Notably, the line-averaged density, $D_\alpha $ emission, and $\beta_N$ remain nearly unchanged throughout this phase, indicating that the plasma stays in L-mode without undergoing an L-H transition, even though the NBI power reaches $1.55 MW$ at $t= 8s$. It is emphasized that the two shots share similar basic plasma parameters, including plasma shaping, heating waveform, and line-averaged density. The primary difference is the application of the ERMP. Therefore, the applied ERMP inhibits the L-H transition, a phenomenon that has also been observed on the HL-3 tokamak \cite{Zhang_2025}.

\begin{figure}[htb!]
	\centering
	\includegraphics[width=\linewidth,scale=1.0]{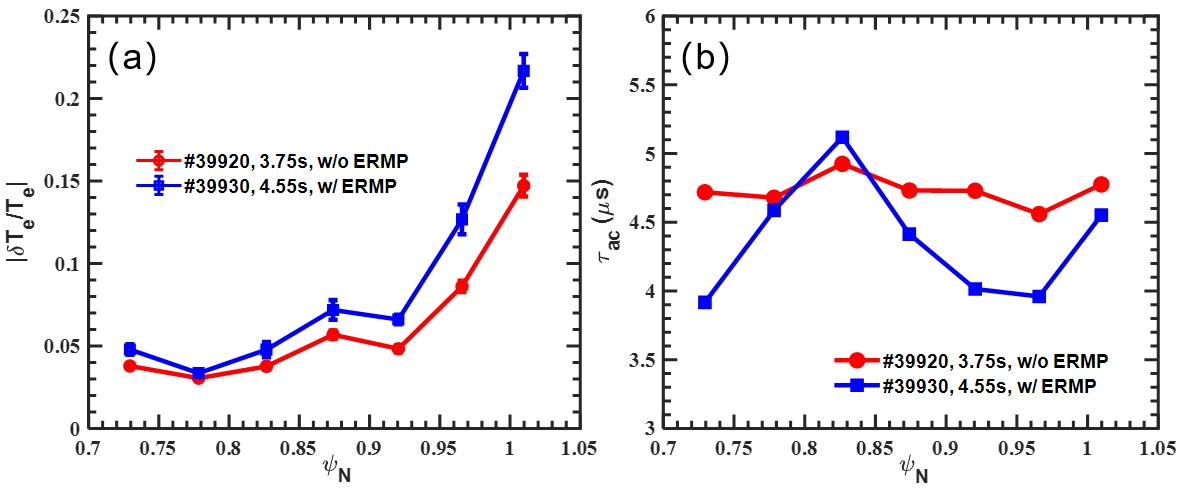}
	\caption{ Radial distribution of (a) electron temperature fluctuations and (b) auto-correlation time in the frequency range of $5-80kHz$, measured by the ECEI diagnostic. The radial coordinate $\psi_N$ is the normalized poloidal magnetic flux and the error bars were estimated by statistical standard deviations during $3.7–3.8 s$ for shot 39920 and $4.5-4.6 s$ for shot 39930, respectively. }
	\label{delta_Te_tau_AC_39920_39930_5-80kHz} 
\end{figure}

\begin{figure}[htb!]
	\centering
	\includegraphics[width=\linewidth,scale=1.0]{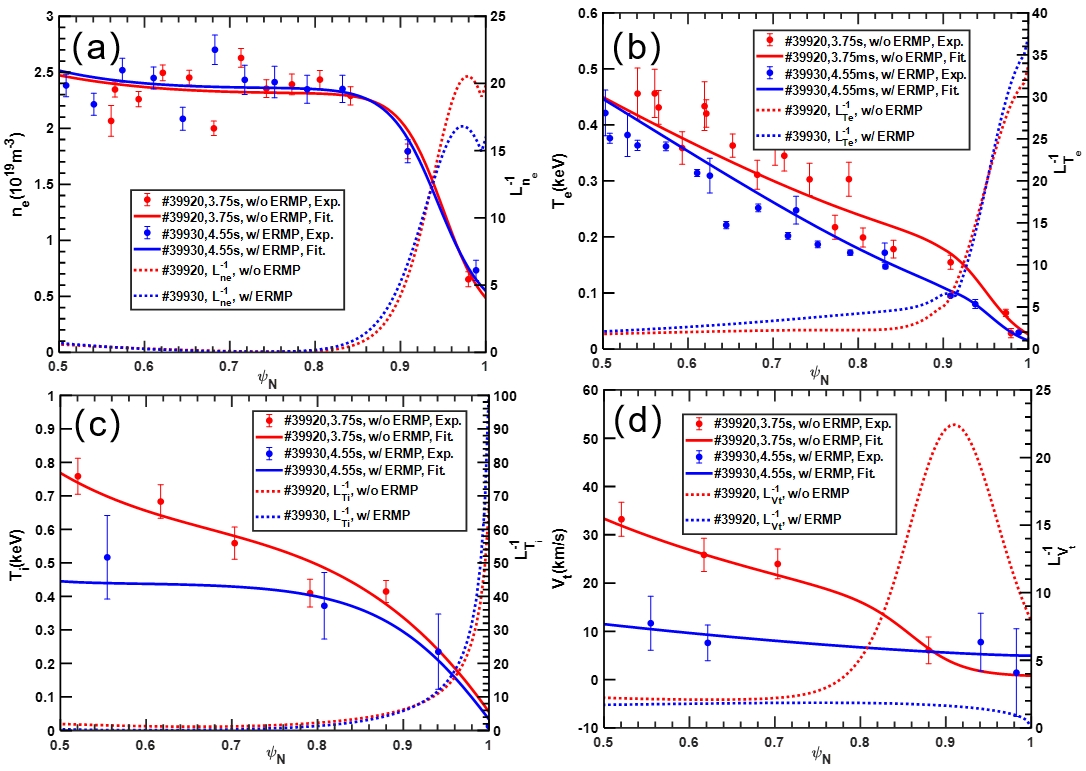}
	\caption{ Equilibrium profiles of (a) electron density $n_e$ and (b) electron temperature $T_e$ measured by the TS diagnostics; (c) ion temperature $T_i$ and (d) toroidal rotation velocity $V_t$ measured by the CES diagnostics. The markers with error bars represent the measured experimental data points and the solid curves with the same color represent corresponding fitting profiles. The dashed curves with the corresponding color represent the related scale length calculated from the fitting curves.}
	\label{Profiles_39920_39930} 
\end{figure}

To understand the mechanism by which the ERMP inhibits the L-H transition, we first examine how the ERMP affects the turbulence characteristics. The comparison is made at two time slices in the L-mode state: one at $t = 3.75 s$ in shot \#39920, just before the L-H transition, and the other one at $t = 4.55 s$ in shot \#39930, which is in a quasi-stationary state with an NBI heating power similar to that in shot \#39920 at $t = 3.75 s$. Figure \ref{delta_Te_tau_AC_39920_39930_5-80kHz} compares the amplitude and autocorrelation time of the electron temperature fluctuations measured by the ECEI diagnostic. With ERMP applied, the electron temperature fluctuation $|\delta T_e/ T_e|$ is enhanced in the edge region of $0.9<\psi_N < 1$ (figure \ref{delta_Te_tau_AC_39920_39930_5-80kHz}(a)). Meanwhile, the autocorrelation time $\tau_{ac}$ decreases from $\sim 4.6\mu s$ to $\sim 4.0\mu s$ in the same region (figure \ref{delta_Te_tau_AC_39920_39930_5-80kHz}(b)). Since the turbulence decorrelation rate can be estimated from the reciprocal of the autocorrelation time, i.e., $\omega_D \approx 1/\tau_{ac}$, the presence of ERMP increases the decorrelation rate from approximately $213kHz$ to $250kHz$.

We then examine how the ERMP affects the equilibrium profiles. Figure \ref{Profiles_39920_39930} compares the equilibrium profiles between shot \#39920 (red markers and solid curves) and shot \#39930 (blue markers and curves). The inverse scale length $L_{f}^{-1} = d\ln f/d\psi_N$, where $f$ denotes the equilibrium quantity, is calculated from the fitted profiles and presented with dotted curves in the figure. With ERMP applied, in the edge region ($\psi_N= 0.8-1$), the electron density remains nearly unchanged, the electron and ion temperature decrease, and the toroidal rotation gets flattened. Regarding the scale lengths in the edge region, the density scale length decreases slightly, the electron and ion temperature scale lengths remain nearly unchanged, and the toroidal rotation scale length decreases substantially. The most pronounced modification of the equilibrium profiles by the ERMP is the flattening of the toroidal rotation and reduction of its gradient in the region of $\psi_N=0.8-1$, which shares a similar trend to that observed on DIII-D tokamak with $n=3$ RMP \cite{Mordijck_2016}. Consequently, the associated $E\times B$ velocity profile and its flow shear are expected to be significantly modified.

\begin{figure}[htb!]
	\centering
	\includegraphics[width=\linewidth,scale=1.0]{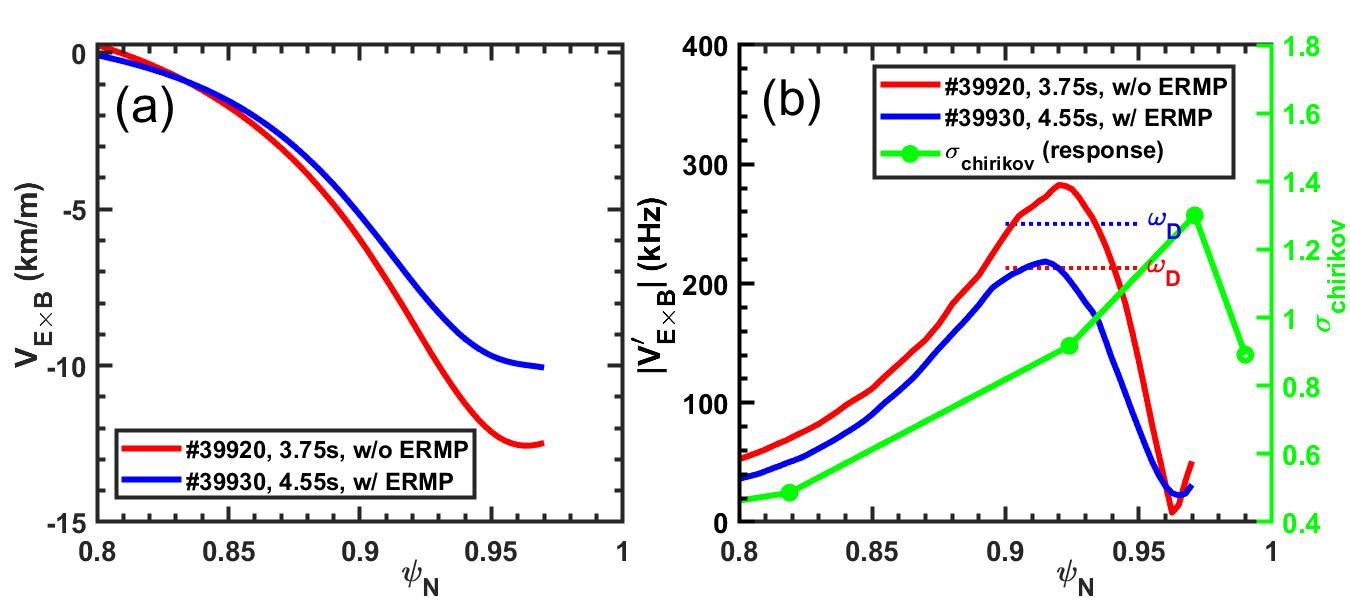}
	\caption{ (a) $E\times B$ velocity $V_{E\times B}$ calculated from the radial force balance. (b) The corresponding $E\times B$ velocity shear with the same color as that in figure \ref{V_ExB_39920_3750ms_39930_4600_Exp}(a). The horizontal dotted lines indicate the decorrelation rate estimated from the auto-correlation time shown in figure \ref{delta_Te_tau_AC_39920_39930_5-80kHz}(b) in the radial range of $0.9<\psi_N<0.95$. The Chirikov parameter $\sigma_{Chirikov}$ profile of shot 39930 is presented in the right y-axis.}
	\label{V_ExB_39920_3750ms_39930_4600_Exp} 
\end{figure}

The $E\times B$ velocity is linked to the radial electric field $E_r$, which can be evaluated from the radial force balance. Its expression is $E_r=\nabla P_i/(Z_ien_i)-V_{p,i}B_t + V_{t,i} B_p$, where the subscript $i$ denotes the ion species, $P_i$ is the ion pressure, $Z_i$ is the ion charge, $n_i$ is the ion density, $e$ is the electron charge, $V_{p,i}$ and $V_{t,i}$ are the poloidal and toroidal velocities, and $B_p$ and $B_t$ are the poloidal and toroidal magnetic fields, respectively. The poloidal velocity $V_{p,i}$ could be estimated from neoclassical predictions \cite{Hinton_1976}. The $E\times B$ velocity is then given by $V_{E\times B} = E_r \times B/B^2$. It should be noted that in the edge region of $0.97<\psi_N<1$, the $E\times B$ velocity obtained from equilibrium diagnostics is subject to large uncertainties because of the limited spatial resolution. Therefore, the present analysis focuses on the inner layer of $\psi_N<0.97$. In addition, taking into account the plasma response, the radial distribution of the Chirikov parameter $\sigma_{\mathrm{Chirikov}}$ \cite{Rechester_1978,Chirikov_1979,Biewer_2003} is calculated as
$\sigma_{\mathrm{Chirikov}} = (W_m + W_{m+1})/\Delta\psi_{m,m+1}$,
where $W_m$ and $W_{m+1}$ are the widths of the magnetic islands for adjacent poloidal mode numbers $m$ and $m+1$, respectively, and $\Delta\psi_{m,m+1}$ is the distance between the corresponding neighboring rational surfaces. When $\sigma_{\mathrm{Chirikov}}>1$, adjacent magnetic islands overlap and magnetic field stochastization occurs. As shown in figure \ref{V_ExB_39920_3750ms_39930_4600_Exp}(b), in shot \#39930, $\sigma_{\mathrm{Chirikov}}$ is close to 1 in the radial range $0.9<\psi_N<0.94$ and exceeds 1 in the range $0.94<\psi_N<0.97$. This indicates that the stochastic magnetic field produced by the ERMP may strongly modify the $E\times B$ profile in $0.9<\psi_N<0.97$ because related theoretical works show that stochastic fields break ambipolarity and generate non-ambipolar radial currents and Maxwell stresses that alter the mean $E\times B$ shear \cite{Guo_2022}.

Figure \ref{V_ExB_39920_3750ms_39930_4600_Exp} compares the $E\times B$ velocity $V_{E\times B}$ and the associated velocity shear $|V_{E\times B}^\prime|$. With ERMP applied, the amplitude of the $E\times B$ velocity is reduced in the edge region of $\psi_N = 0.9-0.97$. In particular, the velocity shear is significantly reduced in the radial range of $\psi_N = 0.9-0.95$. Quantitatively, without ERMP the $E\times B$ shearing rate exceeds the turbulence decorrelation rate estimated from figure \ref{delta_Te_tau_AC_39920_39930_5-80kHz}(b). In contrast, with ERMP the $E\times B$ shearing rate falls below the decorrelation rate. Consequently, the ability of the shear flow to suppress turbulence is weakened. A larger mean flow shear is therefore required to suppress the turbulence and trigger the L-H transition, which in turn necessitates more heating power. This inhibits the transition to H-mode and is consistent with the failure to enter H-mode shown in figure \ref{Diagram_39920_39930_modified} (blue curves). These results demonstrate that the ERMP disrupts the turbulence shear suppression mechanism in the stationary L-mode state preceding the L-H transition.

\section{1D predator-prey simulation in the presence of ERMP \label{Sim}}

\subsection{One-dimensional reduced transport model}
\label{sec:model}

To gain further insight into the physical mechanisms behind these experimental observations, we have extended the five-field reduced mesoscale 1D predator-prey model \cite{Miki_2012,Li_2024,Zhang_2025} to incorporate the toroidal flow evolution and two RMP-related effects: stochastic dephasing of the turbulence-zonal-flow coupling and the resonant Maxwell stress torque. The extended model self-consistently evolves the turbulence intensity $I$, the mean square zonal flow shear $E_{\mathrm{ZF}}=\langle V_{\mathrm{ZF}}'\rangle^{2}$, the ion density $n$, the ion pressure $p$, the mean poloidal mass flow $\langle V_{\theta}\rangle$, and the toroidal flow $\langle V_{\phi}\rangle$ as radially resolved fields in normalized minor radius $r$ and time $t$. The governing equations are

\begin{equation}
	\frac{\partial I}{\partial t} = I \left(\alpha_I \gamma_I - \Delta\omega I - \alpha_{ZF} \frac{1}{1+\lambda_1 \tilde{b}^2} E_{ZF} - \alpha_V E_V \right) + \frac{1}{r} \frac{\partial}{\partial r} \left( r \chi_I \frac{\partial I}{\partial r} \right) \label{eq:turb}
\end{equation}

\begin{equation}
	\frac{\partial E_{ZF}}{\partial t} = \frac{\alpha_{ZF} I E_{ZF}}{1+\zeta_0 E_V} \frac{1}{1+\lambda_1 \tilde{b}^2} - \gamma_{d} E_{ZF} \label{eq:zf}
\end{equation}

\begin{equation}
	\frac{\partial n}{\partial t} = -\frac{1}{r} \frac{\partial}{\partial r} \left[ r \left( -(D_{neo} + D_{turb}) \frac{\partial n}{\partial r} + V_n n \right) \right] + S_n \label{eq:n}
\end{equation}

\begin{equation}
	\frac{\partial p}{\partial t} = -\frac{1}{r} \frac{\partial}{\partial r} \left[ r \left( -(\chi_{neo}  + \chi_{turb}) \frac{\partial p}{\partial r} \right) \right] + Q_p \label{eq:p}
\end{equation}

\begin{equation}
	\frac{\partial \langle V_\theta \rangle}{\partial t} = - \alpha_\theta \gamma_I \frac{\partial I}{\partial r} - \mu_{neo} (\langle V_\theta \rangle - 1.17 c_s \rho_s L_T^{-1}) \label{eq:vtheta}
\end{equation}

\begin{equation}
	\frac{\partial \langle V_\phi \rangle}{\partial t} = \frac{1}{r} \frac{\partial}{\partial r} \left[ r \left(  \chi_\phi \frac{\partial \langle V_\phi \rangle}{\partial r}  - V_{pinch} \langle V_\phi \rangle\right) \right] + \alpha_{\nabla T} L_T^{-1} + S_M + S_{RMP}  \label{eq:vphi}
\end{equation}

The mean \(E\times B\) flow follows from the radial force balance
\begin{equation}
	\langle V_{E\times B}\rangle
	= \frac{1}{Z e B n}\frac{\partial p}{\partial r}
	- \langle V_{\theta}\rangle
	+ \frac{B_{\theta}}{B_{\phi}}\langle V_{\phi}\rangle,
	\label{eq:forcebalance}
\end{equation}
and its shearing rate is evaluated as
\begin{equation}
	\langle V_{E\times B}\rangle'
	= \frac{T}{Z e B_{\phi}}
	\left(\frac{p''}{p}-\frac{p'n'}{n p}\right)
	- \langle V_{\theta}\rangle'
	+ \frac{\partial}{\partial r}
	\left(\frac{B_{\theta}}{B_{\phi}}\langle V_{\phi}\rangle\right),
	\label{eq:shearrate}
\end{equation}
where the toroidal contribution is expanded as
\begin{equation}
	\frac{\partial}{\partial r}
	\left(\frac{B_{\theta}}{B_{\phi}}\langle V_{\phi}\rangle\right)
	= \frac{1}{q R}\frac{\partial}{\partial r}(r\langle V_{\phi}\rangle)
	- \frac{r\langle V_{\phi}\rangle}{q^{2}R}\frac{\partial q}{\partial r}.
	\label{eq:toroidalterm}
\end{equation}
Since the turbulence suppression by flow shear depends on the magnitude rather than the sign of the shear, it enters equations \ref{eq:turb} and \ref{eq:zf} through \(E_V=(\langle V_{E\times B}\rangle')^{2}\).

The physical content of the model is as follows. In equation \ref{eq:turb}, the linear instability drive is represented by \(\alpha_I\gamma_I\), nonlinear self-saturation by \(\Delta\omega I\), zonal-flow suppression by the term proportional to \(E_{\mathrm{ZF}}\), and mean-flow suppression by \(\alpha_V E_V\); the last term describes non-local turbulence spreading. The zonal-flow equation \ref{eq:zf} balances turbulent Reynolds-stress drive against linear damping at rate \(\gamma_d\), with the drive weakened both by mean-flow shear through \((1+\zeta_0 E_V)^{-1}\) and by RMP-induced dephasing through \((1+\lambda_1\tilde{b}^{2})^{-1}\). The density and pressure equations \ref{eq:n} and \ref{eq:p} include neoclassical and turbulent transport, together with Gaussian sources \(S_n\) and \(Q_p\) localized at the edge and core, respectively. The poloidal flow in equation \ref{eq:vtheta} is driven by the turbulent Reynolds stress and relaxed by neoclassical magnetic pumping toward the neoclassical offset. The toroidal flow in equation \ref{eq:vphi} evolves under momentum diffusion, inward pinch, residual stress, an external momentum source \(S_M\) and the resonant Maxwell stress \(S_{\mathrm{RMP}}\).

Two RMP effects are incorporated. First, stochastic dephasing of the turbulence-zonal-flow coupling is described by the factor \(1/(1+\lambda_1\tilde{b}^{2})\), where \(\tilde{b}=\tilde{B}_r/B_{\phi}\) is the normalized radial magnetic perturbation and
\begin{equation}
	\lambda_1 = \frac{q}{2\rho_s^{*2}\sqrt{\beta\epsilon^{2}\epsilon_n}},
	\label{eq:lambda1}
\end{equation}
with \(\rho_s^{*}=\rho_s/a\), \(\beta\) the plasma beta, \(\epsilon=a/R\), and \(\epsilon_n=R/L_n\). Second, the resonant Maxwell stress torque is
\begin{equation}
	S_{\mathrm{RMP}}
	= \frac{1}{n m_i c}\langle J_r\rangle B_{\theta}
	= \frac{c_s^{2}}{\beta}\frac{B_{\theta}}{B}
	\frac{\partial}{\partial r}\langle\tilde{b}_r\tilde{b}_{\theta}\rangle
	\approx -\frac{c_s^{2}}{\beta}\frac{B_{\theta}}{B}V_\phi
	\frac{\partial}{\partial r}\tilde{b}^{2},
	\label{eq:Maxwell}
\end{equation}
where the radial current density \(\langle J_r\rangle=\frac{cB_0}{4\pi}\partial_r\langle\tilde{b}_r\tilde{b}_\theta\rangle\) follows from Ampere's law and the Taylor identity \cite{Guo_2022}. Here \(\tilde{b}\) denotes the self-consistent plasma response field rather than the vacuum perturbation, so that the torque vanishes when the fluctuating parallel current \(\tilde{J}_{||}\to 0\).

The system is normalized by the minor radius \(a\) for the radial coordinate and by the acoustic transit time \(a/c_s\) for time, with \(c_s=\sqrt{T_0/m_i}\). Thermal equilibration \(T_{e0}=T_{i0}=T_0\) is assumed, and all variables and transport coefficients are dimensionless.

\begin{figure}[htb!]
	\centering
	\includegraphics[width=\linewidth,scale=1.0]{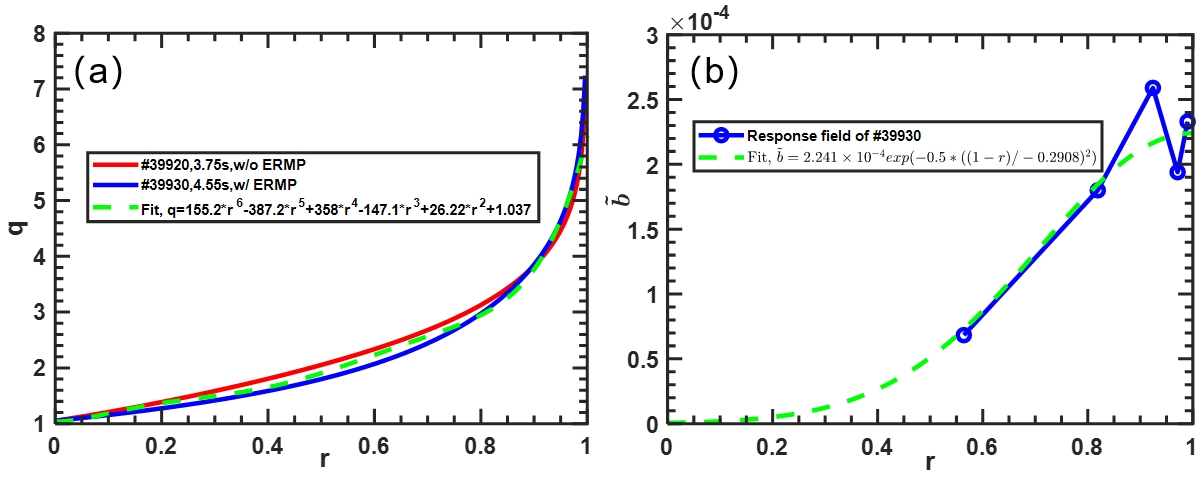}
	\caption{ Radial profiles of (a) safety factor $q$ and (b) perturbed magnetic field $\tilde{b}$ including the plasma response. The green dashed lines indicate the fitted profiles with the fitting formula presented in the legends. }
	\label{q_br_fit_39920_39930} 
\end{figure}

For the KSTAR simulations, the model inputs are calibrated to the experimental conditions of the tungsten-divertor LSN discharges. The reference state adopts $R_0=1.8 m$, $a=0.5 m$, $B_t=1.8 T$. In addition, guided by the experimental profiles, the reference electron density and ion temperature are chosen as typical values near the mid-radius: $n_0 = 2\times 10^{19} m^{-3}$ and $T_0 = 1 keV$. The safety-factor profile is fitted from equilibrium reconstructions of the target discharges, as shown in figure \ref{q_br_fit_39920_39930}(a). The radial distribution of the normalized magnetic perturbation $\tilde{b}$, is obtained by fitting the IPEC code results, which takes into account the plasma response, as shown in figure \ref{q_br_fit_39920_39930}(b). The heating power $Q_p$ is ramped linearly from $3\times 10^{-4}$ to $8.8\times 10^{-3}$ over $t=1\times 10^{4}$ to $t=8 \times 10^{4} $. The remaining input parameters are adopted from previous literature \cite{Miki_2012,Li_2024,Hu_2026s}, without specific tuning.

\subsection{Simulation results}

\begin{figure}[htb!]
	\centering
	\subfigure{
		\label{d7_0.0e-4} 
		\includegraphics[width=0.9\textwidth,scale=1]{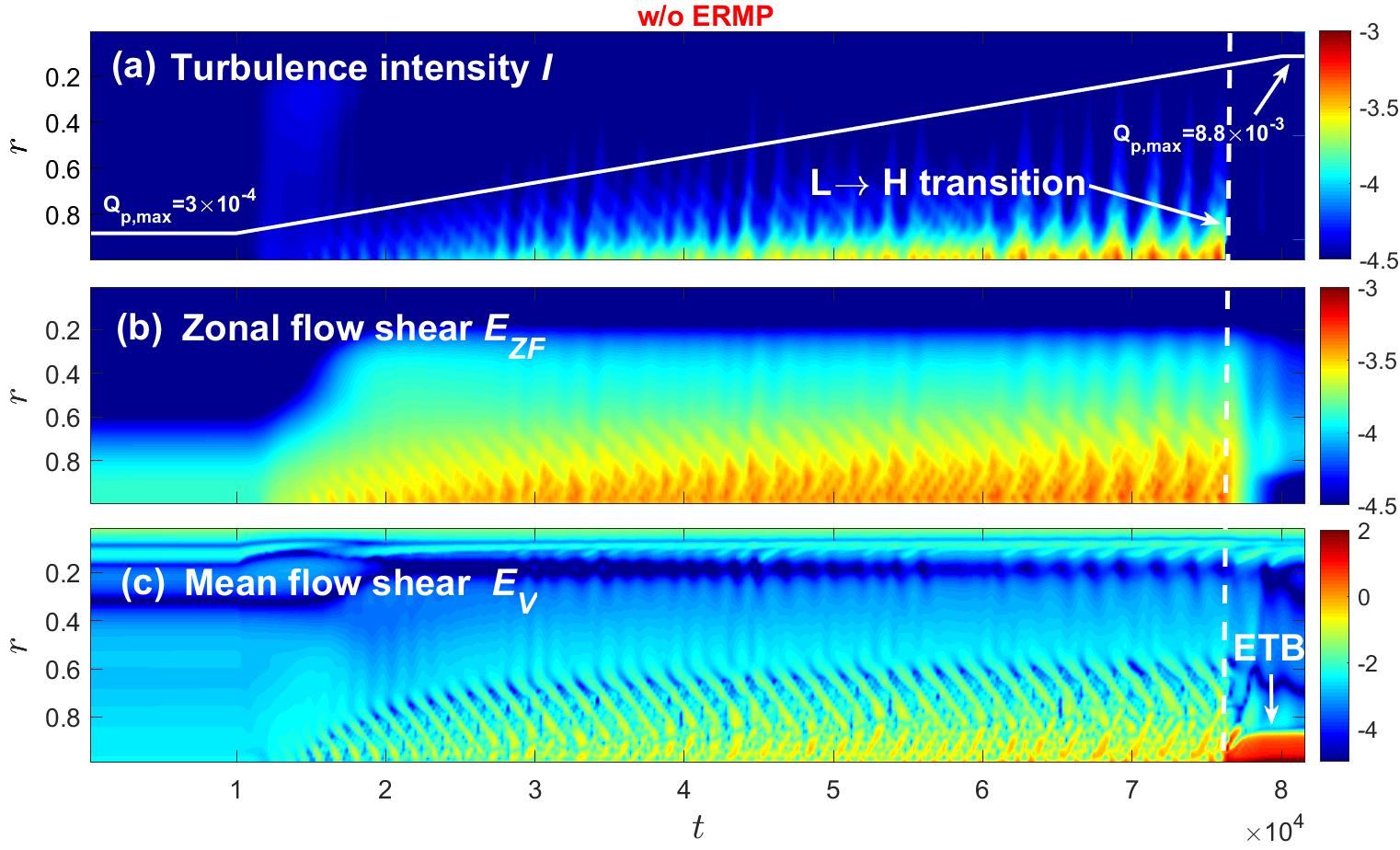}}
	\subfigure{
		\label{d7_2.24e-4} 
		\includegraphics[width=0.9\textwidth,scale=1]{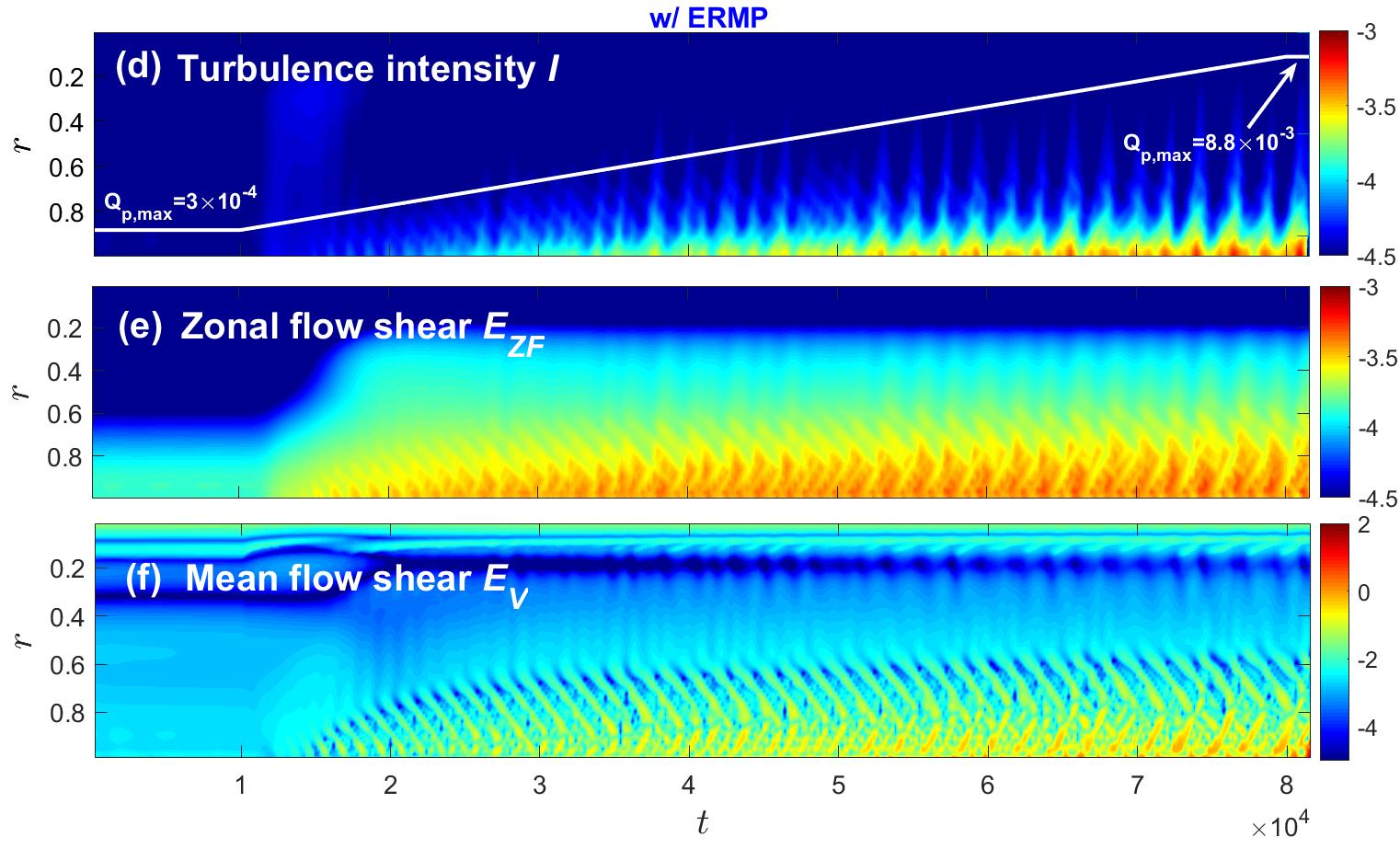}}
	\caption{ Spatiotemporal evolution of turbulence intensity $\log_{10}I$, zonal flow shear $\log_{10}E_{ZF}$ and mean $E\times B$ flow shear $\log_{10}E_V$. (a)-(c) ERMP not included. (d)-(f) ERMP included. The log color scale is chosen so as to easily show the phase evolution. The input heating power is increased from $Q_{p,max} = 3\times 10^{-4}$ to $8.8\times 10^{-3}$, with a duration of $\Delta t = 7\times 10^4$ staring at $t=1\times 10^4$ for the linear power ramp up, as shown with white solid curves in figures (a) and (d). The vertical white dashed line mark the L-H transition time.}
	\label{2D_evolution} 
\end{figure}

Guided by the experimental measurements, the 1D extended predator-prey model is now applied to investigate how the ERMP alters the L-H transition dynamics on KSTAR. All model parameters are held fixed except for the inclusion of the radial magnetic perturbation $\tilde{b}$, which allows the effect of the ERMP to be isolated. The heating power is ramped linearly in time, following the experimental waveform, so that the simulated transition results can be compared with the experimental observations.

Figures \ref{2D_evolution}(a)-(c) present the spatiotemporal evolution of the turbulence intensity $I$, the mean square zonal flow shear $E_{\mathrm{ZF}}$, and the mean $E\times B$ flow shear $E_V$ for the case without ERMP. During the initial phase ($0<t<1\times 10^4$) with low heating power ($Q_{p,\max}=3\times 10^{-4}$), the plasma remains in a quiescent L-mode state in which the turbulence level is low and the zonal flow provides only weak regulation. As the heating power is ramped up from $Q_{p,max} = 3\times 10^{-4}$ to $Q_{p,max} = 8.8\times 10^{-3}$ during  $1\times 10^4 <t< 8\times 10^4$, the system enters the I-phase, characterized by pronounced limit-cycle oscillations (LCOs) of the edge turbulence, zonal flow, and mean flow shear. The LCO phase persists over an extended period, during which the turbulence and zonal flow exhibit periodic bursts and the mean flow shear remains modest. When the heating power exceeds a critical value of $Q_{p,\max}\approx 8.4\times 10^{-3}$ at $t\approx 7.64\times 10^{4}$, the LCO terminates abruptly and the edge turbulence and zonal flow are rapidly quenched. Simultaneously, a strong mean flow shear layer develops at the edge and the edge ETB forms on a fast timescale. The pedestal width saturates shortly afterwards, and a steady H-mode state is maintained for the remainder of the simulation. These features reproduce the characteristic sequence of L-mode, I-phase with LCO, and H-mode that is commonly observed in tokamak experiments.

In contrast, figures \ref{2D_evolution}(d)-(f) show the same quantities for the case with ERMP included and all other parameters unchanged. The plasma again evolves from the initial L-mode into the LCO state as the heating power increases. However, the LCO phase persists for the entire duration of the heating ramp, and no transition to H-mode occurs even when the heating power reaches the maximum value of $Q_{p,\max}=8.8\times 10^{-3}$. The turbulence and zonal flow continue to oscillate quasi-periodically, and the mean flow shear remains below the level required to form a steady ETB. This behavior is in qualitative agreement with the experimental observation that the ERMP inhibits the L-H transition in KSTAR, as shown in figure \ref{Diagram_39920_39930_modified}. The simulation therefore captures the essential effect of the ERMP on the macroscopic transition dynamics.

\begin{figure}[htb!]
	\centering
	\includegraphics[width=\linewidth,scale=1.0]{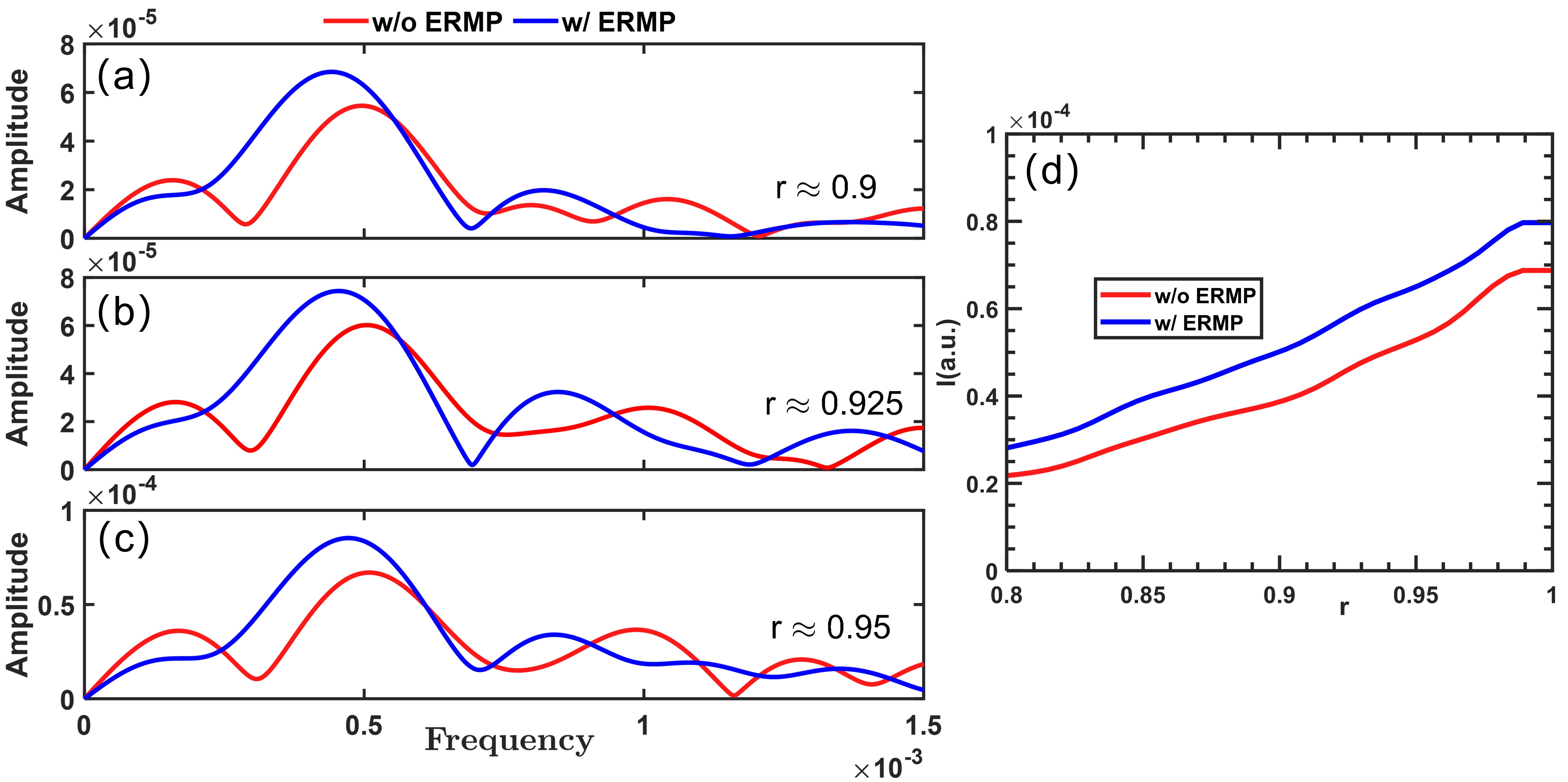}
	\caption{ Comparison of turbulence features in the LCO state during $7.2\times 10 ^{4}<t<7.6\times 10 ^{4}$: (a) power spectra at radial locations of $r=0.9$, $r=0.925$ and $0.95$; (b) the corresponding radial profiles. }
	\label{Sim_I_profile_spectra} 
\end{figure}
	
To quantify the differences between the two cases, figure \ref{Sim_I_profile_spectra} compares the turbulence features in the LCO state over the time window of $7.2\times 10^{4}<t<7.6\times 10^{4}$. The power spectra at $r=0.9$, $r=0.925$, and $r=0.95$ are shown in figures \ref{Sim_I_profile_spectra}(a)-(c). In the absence of ERMP, the spectra exhibit a relatively broad distribution with a modest peak amplitude. With ERMP, the spectral amplitude increases significantly and the spectral weight shifts toward lower frequencies, indicating the excitation of low-frequency coherent modes, which is consistent with the measurements of beam emission spectroscopy on the DIII-D tokamak \cite{Schmitz_2019}. Figure \ref{Sim_I_profile_spectra}(d) compares the corresponding radial distribution of the turbulence intensity. It shows that, with ERMP, the turbulence intensity is enhanced over a broad radial range. This trend is consistent with the experimental ECEI measurements shown in figure \ref{delta_Te_tau_AC_39920_39930_5-80kHz}, where the ERMP enhances the electron temperature fluctuations in the edge region.

The underlying reason for the turbulence enhancement can be understood from the model equations. The ERMP introduces a dephasing factor $1/(1+\lambda_1\tilde{b}^{2})$ into the turbulence and zonal-flow equations, which weakens the energy transfer from turbulence to zonal flows. As a result, the zonal flow is less effective in regulating the turbulence, and the turbulence saturates at a higher level. In addition, the Maxwell stress torques modify the toroidal rotation profile and reduce the mean flow shear, further degrading the turbulence suppression. These two effects act in concert to delay the buildup of the edge flow shear and to raise the heating power required for the L-H transition.

\begin{figure}[htb!]
	\centering
	\includegraphics[width=\linewidth,scale=1.0]{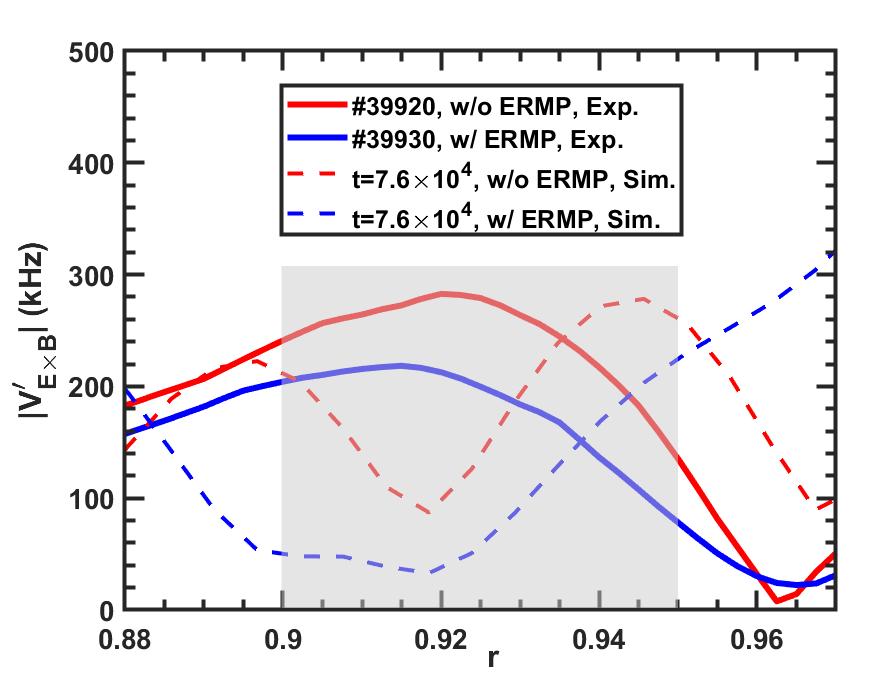}
	\caption{ Comparison of $E\times B$ shearing rate in the LCO state at $t=7.6\times 10 ^{4}$. The solid lines represent the experimental results of KSTAR shown in figure \ref{V_ExB_39920_3750ms_39930_4600_Exp}(b), and the dashed lines represent the related simulation results, respectively. The red/blue lines represent the case without/with ERMP. The grey shaded region marks the shear layer where the ERMP reduces the $E\times B$ shearing rate significantly.}
	\label{V_ExB_39920_3750ms_39930_4600_sim} 
\end{figure}

To make a more direct connection with the experiment, figure \ref{V_ExB_39920_3750ms_39930_4600_sim} compares the simulated $E\times B$ shearing rate with the experimental results shown in figure \ref{V_ExB_39920_3750ms_39930_4600_Exp}(b). The solid lines represent the experimental data, while the dashed lines represent the simulation results, with red and blue denoting the cases without and with ERMP, respectively. In the simulation, the presence of ERMP reduces the $E\times B$ shearing rate in the edge radial range of $0.9<r<0.95$ (gred shaded region in figure \ref{V_ExB_39920_3750ms_39930_4600_sim}), in agreement with the experimental trend. The radial distribution of the shearing rate is qualitatively similar to the experimental observation, and the magnitude of the $E\times B$ shearing rate is quantitatively close to the measured value. This modification of the $E\times B$ shearing rate is the key condition that determines whether the shear flow can suppress the turbulence and trigger the L-H transition.

Taken together, the simulation results demonstrate that the extended 1D model captures the essential physics of ERMP-inhibited L-H transition on KSTAR. The model reproduces the experimental trends in the turbulence amplitude and the edge $E\times B$ shearing rate, and it provides a consistent explanation for the observed delay of the L-H transition. These results also highlight the importance of incorporating the plasma response into the radial magnetic field perturbation $\tilde{b}$, since the radial distribution of $\tilde{b}$ determines the spatial range over which the dephasing and torque effects operate.

\section{Conclusion and discussion \label{Con}}

In this work, we have investigated the impact of ERMP on the L-H transition in the KSTAR discharges with a tungsten LSN divertor, combining dedicated experiments with a one-dimensional reduced transport model. The main findings are summarized as follows.

Experimentally, under nearly identical plasma parameters including shaping, heating waveform, and line-averaged density, the application of an $n=1$ ERMP with a coil current of $1.5 kAt$ inhibits the L-H transition. In the reference discharge without ERMP, the L-H transition occurs at $P_{\mathrm{NBI}}=1.2 MW$, as indicated by the rapid increase of line-averaged density and $\beta_N$ together with a sudden drop of the $D_\alpha$ signal. In contrast, with ERMP applied, the plasma remains in L-mode even when the NBI power reaches $1.55 MW$. This observation is consistent with previous findings on HL-3 \cite{Zhang_2025}, and it provides direct evidence in a tungsten divertor environment.

The mechanism behind the inhibited transition is revealed by the edge turbulence and flow measurements. The ECEI diagnostic shows that the ERMP enhances the electron temperature fluctuation $|\delta T_e/T_e|$ and reduces the autocorrelation time $\tau_{ac}$ in the edge region $0.9<\psi_N<1$, corresponding to an increase of the turbulence decorrelation rate. The equilibrium profiles are also modified: the toroidal rotation becomes flattened across the edge range, while the electron density and ion temperature change modestly. Consequently, the $E\times B$ velocity and its flow shear are reduced in the edge region, and the $E\times B$ shearing rate falls below the turbulence decorrelation rate. This weakening of the shear suppression mechanism necessitates a higher heating power to increase the mean flow shear and thus to trigger the L-H transition, thereby explaining the experimental observation.

The extended 1D predator-prey model, which incorporates the toroidal flow evolution and two RMP-related effects (stochastic dephasing of the turbulence-zonal-flow coupling and the resonant Maxwell stress torque) reproduces the experimental trends. With the radial profile of the normalized magnetic field perturbation obtained from IPEC calculations that include the plasma response, the simulation shows that the ERMP inhibites the L-H transition and enhances the turbulence intensity in the LCO state. The simulated $E\times B$ shearing rate is reduced in the radial range of $0.9<r<0.95$ and is quantitatively close to the experimental values. While the turbulence intensity enhances, consistent with the ECEI measurements. These results demonstrate that the model captures the essential physics of the ERMP-induced inhibition of the L-H transition in KSTAR. The agreement between experiment and simulation supports the conclusion that the fundamental mechanism by which the ERMP affects the L-H transition is similar across different tokamaks and divertor materials. The tungsten divertor environment, while introducing different impurity and recycling dynamics, does not alter this basic picture.

Further experimental and modeling efforts are needed to deepen our understanding. On the experimental side, systematic scans of the RMP intensity and toroidal mode number on both KSTAR and HL-3 would help to clarify how the transition dynamics depend on the perturbation spectrum and the plasma response. In particular, the critical magnetic field perturbation threshold above which the L-H power threshold increases should be determined for different divertor materials. On the modeling side, the 1D model can be improved by incorporating two-fluid and electromagnetic effects, which may become important in the high-beta H-mode regime. These improvements will enable more quantitative predictions for future devices such as ITER, where reliable control of the L-H transition and ELMs is essential.

\ack

We acknowledge Longwen Yan, Neng Zhang, Jie Huang and Lian Wang for valuable discussions. This project was supported by the National Key R\&D Program of China (Nos. 2022YFE03060003 and 2024YFE03270400), the National Natural Science Foundation of China (Nos. 12575236, 12575231, 12405256, 12375210 and 12435014), the Nuclear technology research and development program (No. HJSYF2024(02)), the Sichuan Science and Technology Program (No. 2026NSFSCZY0059) and the Innovation Program of Southwestern Institute of Physics (No. 202301XWCX001-02). This research was also supported by R\&D Program of ``High Performance Tokamak Plasma Research \& Development (code No. EN2601)'' through the Korea Institute of Fusion Energy (KFE) funded by the Government funds, Republic of Korea.

\section*{References}
\bibliography{iopart-num}

@article{Wagner_1982,
  title = {Regime of Improved Confinement and High Beta in Neutral-Beam-Heated Divertor Discharges of the ASDEX Tokamak},
  author = {Wagner, F. and Becker, G. and Behringer, K. and Campbell, D. and Eberhagen, A. and Engelhardt, W. and Fussmann, G. and Gehre, O. and Gernhardt, J. and Gierke, G. v. and Haas, G. and Huang, M. and Karger, F. and Keilhacker, M. and Kl\"uber, O. and Kornherr, M. and Lackner, K. and Lisitano, G. and Lister, G. G. and Mayer, H. M. and Meisel, D. and M\"uller, E. R. and Murmann, H. and Niedermeyer, H. and Poschenrieder, W. and Rapp, H. and R\"ohr, H. and Schneider, F. and Siller, G. and Speth, E. and St\"abler, A. and Steuer, K. H. and Venus, G. and Vollmer, O. and Y\"u, Z.},
  journal = {Phys. Rev. Lett.},
  volume = {49},
  issue = {19},
  pages = {1408--1412},
  numpages = {0},
  year = {1982},
  month = {Nov},
  publisher = {American Physical Society},
  doi = {10.1103/PhysRevLett.49.1408},
  url = {https://link.aps.org/doi/10.1103/PhysRevLett.49.1408}
}

@article{Leonard_2014,
author = {Leonard,A. W. },
title = {Edge-localized-modes in tokamaks},
journal = {Physics of Plasmas},
volume = {21},
number = {9},
pages = {090501},
year = {2014},
doi = {10.1063/1.4894742},
URL = {https://doi.org/10.1063/1.4894742}
}

@article{Wagner_1984,
  title = {Development of an Edge Transport Barrier at the H-Mode Transition of ASDEX},
  author = {Wagner, F. and Fussmann, G. and Grave, T. and Keilhacker, M. and Kornherr, M. and Lackner, K. and McCormick, K. and M\"uller, E. R. and St\"abler, A. and Becker, G. and Bernhardi, K. and Ditte, U. and Eberhagen, A. and Gehre, O. and Gernhardt, J. and Gierke, G. v. and Glock, E. and Gruber, O. and Haas, G. and Hesse, M. and Janeschitz, G. and Karger, F. and Kissel, S. and Kl\"uber, O. and Lisitano, G. and Mayer, H. M. and Meisel, D. and Mertens, V. and Murmann, H. and Poschenrieder, W. and Rapp, H. and R\"ohr, H. and Ryter, F. and Schneider, F. and Siller, G. and Smeulders, P. and S\"oldner, F. and Speth, E. and Steuer, K. -H. and Szymanski, Z. and Vollmer, O.},
  journal = {Phys. Rev. Lett.},
  volume = {53},
  issue = {15},
  pages = {1453--1456},
  numpages = {0},
  year = {1984},
  month = {Oct},
  publisher = {American Physical Society},
  doi = {10.1103/PhysRevLett.53.1453},
  url = {https://link.aps.org/doi/10.1103/PhysRevLett.53.1453}
}

@article{Evans_2004,
  title = {Suppression of Large Edge-Localized Modes in High-Confinement DIII-D Plasmas with a Stochastic Magnetic Boundary},
  author = {Evans, T. E. and Moyer, R. A. and Thomas, P. R. and Watkins, J. G. and Osborne, T. H. and Boedo, J. A. and Doyle, E. J. and Fenstermacher, M. E. and Finken, K. H. and Groebner, R. J. and Groth, M. and Harris, J. H. and La Haye, R. J. and Lasnier, C. J. and Masuzaki, S. and Ohyabu, N. and Pretty, D. G. and Rhodes, T. L. and Reimerdes, H. and Rudakov, D. L. and Schaffer, M. J. and Wang, G. and Zeng, L.},
  journal = {Phys. Rev. Lett.},
  volume = {92},
  issue = {23},
  pages = {235003},
  numpages = {4},
  year = {2004},
  month = {Jun},
  publisher = {American Physical Society},
  doi = {10.1103/PhysRevLett.92.235003}
}

@article{Jeon_2012,
  title = {Suppression of Edge Localized Modes in High-Confinement KSTAR Plasmas by Nonaxisymmetric Magnetic Perturbations},
  author = {Jeon, Y. M. and Park, J.-K. and Yoon, S. W. and Ko, W. H. and Lee, S. G. and Lee, K. D. and Yun, G. S. and Nam, Y. U. and Kim, W. C. and Kwak, Jong-Gu and Lee, K. S. and Kim, H. K. and Yang, H. L.},
  collaboration = {KSTAR team},
  journal = {Phys. Rev. Lett.},
  volume = {109},
  issue = {3},
  pages = {035004},
  numpages = {5},
  year = {2012},
  month = {Jul},
  publisher = {American Physical Society},
  doi = {10.1103/PhysRevLett.109.035004}
}

@article{Sun_2016,
  title = {Nonlinear Transition from Mitigation to Suppression of the Edge Localized Mode with Resonant Magnetic Perturbations in the EAST Tokamak},
  author = {Sun, Y. and Liang, Y. and Liu, Y. Q. and Gu, S. and Yang, X. and Guo, W. and Shi, T. and Jia, M. and Wang, L. and Lyu, B. and Zhou, C. and Liu, A. and Zang, Q. and Liu, H. and Chu, N. and Wang, H. H. and Zhang, T. and Qian, J. and Xu, L. and He, K. and Chen, D. and Shen, B. and Gong, X. and Ji, X. and Wang, S. and Qi, M. and Song, Y. and Yuan, Q. and Sheng, Z. and Gao, G. and Fu, P. and Wan, B.},
  journal = {Phys. Rev. Lett.},
  volume = {117},
  issue = {11},
  pages = {115001},
  numpages = {5},
  year = {2016},
  month = {Sep},
  publisher = {American Physical Society},
  doi = {10.1103/PhysRevLett.117.115001}
}

@article{Sun_2021,
doi = {10.1088/1741-4326/abd2c7},
url = {https://dx.doi.org/10.1088/1741-4326/abd2c7},
year = {2021},
month = {feb},
publisher = {IOP Publishing},
volume = {61},
number = {3},
pages = {036020},
author = {T.F. Sun and Y. Liu and X.Q. Ji and Y.Q. Liu and R. Ke and J.M. Gao and N. Wu and W. Deng and M. Xu and X.R. Duan},
title = {Edge-coherent oscillation providing nearly continuous transport during edge-localized mode mitigation by n = 1 resonant magnetic perturbation in HL-2A},
journal = {Nuclear Fusion}
}

@article{Ryter_2012,
doi = {10.1088/0029-5515/52/11/114014},
url = {https://dx.doi.org/10.1088/0029-5515/52/11/114014},
year = {2012},
month = {oct},
publisher = {IOP Publishing and International Atomic Energy Agency},
volume = {52},
number = {11},
pages = {114014},
author = {F. Ryter and S.K. Rathgeber and E. Viezzer and W. Suttrop and A. Burckhart and R. Fischer and B. Kurzan and S. Potzel and T. Pütterich and the ASDEX Upgrade Team},
title = {L–H transition in the presence of magnetic perturbations in ASDEX Upgrade},
journal = {Nuclear Fusion}
}

@article{In_2017,
doi = {10.1088/1741-4326/aa791c},
url = {https://dx.doi.org/10.1088/1741-4326/aa791c},
year = {2017},
month = {aug},
publisher = {IOP Publishing},
volume = {57},
number = {11},
pages = {116054},
author = {Y. In and J. K. Park and Y. M. Jeon and J. Kim and G.Y. Park and J.W. Ahn and A. Loarte and W.H. Ko and H.H. Lee and J.W. Yoo and J.W. Juhn and S.W. Yoon and H. Park},
title = {Enhanced understanding of non-axisymmetric intrinsic and controlled field impacts in tokamaks},
journal = {Nuclear Fusion}
}

@article{Leonard_1991,
doi = {10.1088/0029-5515/31/8/008},
url = {https://dx.doi.org/10.1088/0029-5515/31/8/008},
year = {1991},
month = {aug},
publisher = {},
volume = {31},
number = {8},
pages = {1511},
author = {A.W. Leonard and A.M. Howald and A.W. Hyatt and T. Shoji and T. Fujita and M. Miura and N. Suzuki and S. Tsuji and JFT-2M Group},
title = {Effects of applied error fields on the H-mode power threshold of JFT-2M},
journal = {Nuclear Fusion}
}

@article{Gohil_2011,
doi = {10.1088/0029-5515/51/10/103020},
url = {https://dx.doi.org/10.1088/0029-5515/51/10/103020},
year = {2011},
month = {aug},
publisher = {},
volume = {51},
number = {10},
pages = {103020},
author = {P. Gohil and T.E. Evans and M.E. Fenstermacher and J.R. Ferron and T.H. Osborne and J.M. Park and O. Schmitz and J.T. Scoville and E.A. Unterberg},
title = {L–H transition studies on DIII-D to determine H-mode access for operational scenarios in ITER},
journal = {Nuclear Fusion}
}

@article{Suttrop_2017,
doi = {10.1088/0741-3335/59/1/014049},
url = {https://dx.doi.org/10.1088/0741-3335/59/1/014049},
year = {2016},
month = {nov},
publisher = {IOP Publishing},
volume = {59},
number = {1},
pages = {014049},
author = {W Suttrop and A Kirk and R Nazikian and N Leuthold and E Strumberger and M Willensdorfer and M Cavedon and M Dunne and R Fischer and S Fietz and J C Fuchs and Y Q Liu and R M McDermott and F Orain and D A Ryan and E Viezzer and The ASDEX Upgrade Team and The DIII-D Team and The Eurofusion MST1 Team},
title = {Experimental studies of high-confinement mode plasma response to non-axisymmetric magnetic perturbations in ASDEX Upgrade},
journal = {Plasma Physics and Controlled Fusion}
}

@article{Schmitz_2019,
doi = {10.1088/1741-4326/ab36bf},
url = {https://dx.doi.org/10.1088/1741-4326/ab36bf},
year = {2019},
month = {sep},
publisher = {IOP Publishing},
volume = {59},
number = {12},
pages = {126010},
author = {L. Schmitz and D.M. Kriete and R.S. Wilcox and T.L. Rhodes and L. Zeng and Z. Yan and G.R. McKee and T.E. Evans and C. Paz-Soldan and P. Gohil and B. Lyons and C.C. Petty and D. Orlov and A. Marinoni},
title = {L–H transition trigger physics in ITER-similar plasmas with applied n  =  3 magnetic perturbations},
journal = {Nuclear Fusion}
}

@article{Kriete_2020,
    author = {Kriete, D. M. and McKee, G. R. and Schmitz, L. and Smith, D. R. and Yan, Z. and Morton, L. A. and Fonck, R. J.},
    title = {Effect of magnetic perturbations on turbulence-flow dynamics at the L-H transition on DIII-D},
    journal = {Physics of Plasmas},
    volume = {27},
    number = {6},
    pages = {062507},
    year = {2020},
    month = {06},
    issn = {1070-664X},
    doi = {10.1063/1.5145207},
    url = {https://doi.org/10.1063/1.5145207}
}

@article{Kirk_2011,
doi = {10.1088/0741-3335/53/6/065011},
url = {https://dx.doi.org/10.1088/0741-3335/53/6/065011},
year = {2011},
month = {apr},
publisher = {},
volume = {53},
number = {6},
pages = {065011},
author = {A Kirk and Yueqiang Liu and E Nardon and P Tamain and P Cahyna and I Chapman and P Denner and H Meyer and S Mordijck and D Temple and the MAST team},
title = {Magnetic perturbation experiments on MAST L- and H-mode plasmas using internal coils},
journal = {Plasma Physics and Controlled Fusion}
}

@article{Kaye_2011,
doi = {10.1088/0029-5515/51/11/113019},
url = {https://dx.doi.org/10.1088/0029-5515/51/11/113019},
year = {2011},
month = {oct},
publisher = {},
volume = {51},
number = {11},
pages = {113019},
author = {S.M. Kaye and R. Maingi and D. Battaglia and R.E. Bell and C.S. Chang and J. Hosea and H. Kugel and B.P. LeBlanc and H. Meyer and G.Y. Park and J.R. Wilson},
title = {L–H threshold studies in NSTX},
journal = {Nuclear Fusion}
}

@article{Yang_2024,
  title={Tailoring tokamak error fields to control plasma instabilities and transport},
  author={Yang, SeongMoo and Park, Jong-Kyu and Jeon, YoungMu and Logan, Nikolas C and Lee, Jaehyun and Hu, Qiming and Lee, JongHa and Kim, SangKyeun and Kim, Jaewook and Lee, Hyungho and others},
  journal={Nature Communications},
  volume={15},
  number={1},
  pages={1275},
  year={2024},
  doi = {10.1038/s41467-024-45454-1},
  url = {https://doi.org/10.1038/s41467-024-45454-1},
  publisher={Nature Publishing Group UK London}
}

@article{Wang_2024,
title = {Final development and preliminary experiment progress of in-vessel resonant magnetic perturbation coils system on HL-3 tokamak},
journal = {Fusion Engineering and Design},
volume = {208},
pages = {114702},
year = {2024},
issn = {0920-3796},
doi = {https://doi.org/10.1016/j.fusengdes.2024.114702},
url = {https://www.sciencedirect.com/science/article/pii/S0920379624005520},
author = {A. Wang and T.F. Sun and W. Chen and B.T. Cui and J.M. Gao and S.Y. Liang and M.Y. He and X.Q. Ji}
}

@article{Chen_2021,
    author = {Chen, Chang-Chun and Diamond, Patrick H. and Singh, Rameswar and Tobias, Steven M.},
    title = {Potential vorticity transport in weakly and strongly magnetized plasmas},
    journal = {Physics of Plasmas},
    volume = {28},
    number = {4},
    pages = {042301},
    year = {2021},
    month = {04},
    issn = {1070-664X},
    doi = {10.1063/5.0041072},
    url = {https://doi.org/10.1063/5.0041072},
}

@article{Li_2024,
doi = {10.1088/1741-4326/ad2d39},
url = {https://dx.doi.org/10.1088/1741-4326/ad2d39},
year = {2024},
month = {mar},
publisher = {IOP Publishing},
volume = {64},
number = {4},
pages = {044001},
author = {Xiaobo Li and Zhibin Guo and Patrick H. Diamond},
title = {Formation, propagation and conversion of transport barriers triggered by dynamical critical gradient in tokamak plasmas},
journal = {Nuclear Fusion},
}

@article{Miki_2012,
    author = {Miki, K. and Diamond, P. H. and Gurcan, O. D. and Tynan, G. R. and Estrada, T. and Schmitz, L. and Xu, G. S.},
    title = {Spatio-temporal evolution of the L → I → H transition},
    journal = {Physics of Plasmas},
    volume = {19},
    number = {9},
    pages = {092306},
    year = {2012},
    month = {09},
    issn = {1070-664X},
    doi = {10.1063/1.4753931},
    url = {https://doi.org/10.1063/1.4753931},
}

@article{Willensdorfer_2022,
    author = {Willensdorfer, M. and Plank, U. and Brida, D. and Cavedon, M. and Conway, G. D. and Ryan, D. A. and Suttrop, W. and Buchholz, R. and Dunne, M. and Fischer, R. and Griener, M. and Hobirk, J. and Kasilov, S. and Kirk, A. and McDermott, R. M. and Pütterich, T. and Tardini, G. and Yu, Q. and ASDEX Upgrade Team},
    title = {Dependence of the L–H power threshold on the alignment of external non-axisymmetric magnetic perturbations in ASDEX Upgrade},
    journal = {Physics of Plasmas},
    volume = {29},
    number = {3},
    pages = {032506},
    year = {2022},
    month = {03},
    issn = {1070-664X},
    doi = {10.1063/5.0073841},
    url = {https://doi.org/10.1063/5.0073841}
}

@article{Leconte_2014,
doi = {10.1088/0029-5515/54/1/013004},
url = {https://dx.doi.org/10.1088/0029-5515/54/1/013004},
year = {2013},
month = {nov},
publisher = {IOP Publishing},
volume = {54},
number = {1},
pages = {013004},
author = {Leconte, M. and Diamond, P.H. and Xu, Y.},
title = {Impact of resonant magnetic perturbations on zonal modes, drift-wave turbulence and the L–H transition threshold},
journal = {Nuclear Fusion}
}

@article{Zhang_2025,
doi = {10.1088/1741-4326/add35b},
url = {https://doi.org/10.1088/1741-4326/add35b},
year = {2025},
month = {may},
publisher = {IOP Publishing},
volume = {65},
number = {6},
pages = {066016},
author = {Zhang, Y. and Jiang, M. and Guo, Z.B. and Diamond, P.H. and Guo, W.X. and Xu, J.Q. and Zhang, N. and Wang, Z. and Li, Z.J. and Zhou, Y. and Liang, A.S. and Sun, T.F. and Ke, R. and Gong, S.B. and He, X.X. and Chen, W. and Zhong, W.L.},
title = {Influence of n = 1 resonant magnetic perturbation on flow and turbulence towards L-H transition},
journal = {Nuclear Fusion}
}

@article{Park_2019,
doi = {10.1088/1741-4326/ab20e2},
url = {https://doi.org/10.1088/1741-4326/ab20e2},
year = {2019},
month = {jul},
publisher = {IOP Publishing},
volume = {59},
number = {11},
pages = {112020},
author = {Park, H.K. and Choi, M.J. and Hong, S.H. and In, Y. and Jeon, Y.M. and Ko, J.S. and Ko, W.H. and Kwak, J.G. and Kwon, J.M. and Lee, J. and Lee, J.H. and Lee, W. and Nam, Y.B. and Oh, Y.K. and Park, B.H. and Park, J.K. and Park, Y.S. and Wang, S.J. and Yoo, M. and Yoon, S.W. and Bak, J.G. and Chang, C.S. and Choe, W.H. and Chu, Y. and Chung, J. and Eidietis, N. and Han, H.S. and Hahn, S.H. and Jhang, H.G. and Juhn, J.W. and Kim, J.H. and Kim, K. and Loarte, A. and Lee, H.H. and Lee, K.C. and Mueller, D. and Na, Y.S. and Nam, Y.U. and Park, G.Y. and Park, K.R. and Pitts, R.A. and Sabbagh, S.A. and Yun, G.S. and the KSTAR Team},
title = {Overview of KSTAR research progress and future plans toward ITER and K-DEMO},
journal = {Nuclear Fusion}
}

@article{Diamond_2013,
  title = {An overview of intrinsic torque and momentum
transport bifurcations in toroidal plasmas},
  author = {Diamond, P.H. and Kosuga, Y. and G\"urcan, \"O.D. and McDevitt, C.J. and Hahm, T.S. and Fedorczak, N. and Rice, J.E. and Wang, W.X. and Ku, S. and Kwon, J.M. and others},
  journal = {Nuclear Fusion},
  volume = {53},
  number = {10},
  pages = {104019},
  year = {2013},
  publisher = {IOP Publishing},
  doi = {10.1088/0029-5515/53/10/104019},
  url = {https://doi.org/10.1088/0029-5515/53/10/104019}
}

@article{Miki_2013,
doi = {10.1088/0029-5515/53/7/073044},
url = {https://doi.org/10.1088/0029-5515/53/7/073044},
year = {2013},
month = {jun},
publisher = {IOP Publishing and International Atomic Energy Agency},
volume = {53},
number = {7},
pages = {073044},
author = {Miki, K. and Diamond, P.H. and Fedorczak, N. and Gurcan, O.D. and Malkov, M. and Lee, C. and Kosuga, Y. and Tynan, G. and Xu, G.S. and Estrada, T. and McDonald, D. and Schmitz, L. and Zhao, K.J.},
title = {Spatio-temporal evolution of the L-H and H-L transitions},
journal = {Nuclear Fusion}
}

@unpublished{Hu_2026s,
  author = {Hu, J. D. and Zhang, Y. and Jiang, M. and others},
  title = {Impact of resonant magnetic perturbations on {L}-{H} and {H}-{L} transition dynamics in tokamak plasmas. },
  note = {Submitted to \textit{Nuclear Fusion}},
  year = {2026}
}

@unpublished{Kim_2025_APS,
  author = {Kim, Eun-Jin},
  title = {The Role of Edge Resonant Magnetic Perturbations in {KSTAR} {L}-{H} Transitions: A Statistical and Geometric View. },
  note = {{APS} Division of Plasma Physics Meeting 2025, Oral presentation, PO05.00012},
  year = {2025},
  url = {https://meetings-archive.aps.org/dpp/2025/po05/12/}
}

@unpublished{Kim_2025_IAEA,
  author = {Kim, E. and Choi, M. J. and Han, S. J. and Yang, S. M. and Park, J.-G. and Ko, W. H. and Juhn, J. W. and Seo, J. H. and Kim, H. S. and KSTAR Team},
  title = {Novel Effects of Edge-Localised {RMPs} and Plasma Density on the {L}-{H} Transitions and Transport. },
  note = {30th IAEA Fusion Energy Conference (IAEA FEC 2025), EX-P},
  year = {2025},
  url = {https://conferences.iaea.org/event/392/contributions/36310/}
}

@article{Kim_2025,
  author = {Kim, Eun-jin and Chung, Huiwon and Choi, Minjun and Lee, Yoonji and Han, Seongjun and Choi, Gyungjin and Park, Jong Gyu and Yang, Seong Moo and Na, Yong-Su},
  title = {Turbulence statistical analysis of the {L}-{H} transition and {RMPs} in {KSTAR}},
  journal = {Plasma Physics and Controlled Fusion},
  volume = {67},
  number = {10},
  pages = {105002},
  year = {2025},
  doi = {10.1088/1361-6587/ae087a},
  url = {https://doi.org/10.1088/1361-6587/ae087a}
}

@article{Kim_2025TS,
doi = {10.1088/1748-0221/20/10/C10003},
url = {https://doi.org/10.1088/1748-0221/20/10/C10003},
year = {2025},
month = {oct},
publisher = {IOP Publishing},
volume = {20},
number = {10},
pages = {C10003},
author = {Kim, Jung-Hwa and Lee, Jong-ha and Kim, Young-Gi and Kim, Hajin and Park, Geunhyeong},
title = {Relative calibration correction using fast LEDs to enhance the reliability of Thomson scattering diagnostic on KSTAR},
journal = {Journal of Instrumentation}
}

@article{Ko_2010,
    author = {Ko, Won-Ha and Lee, Hyungho and Seo, Dongcheol and Kwon, Myeun},
    title = {Charge exchange spectroscopy system calibration for ion temperature measurement in KSTAR},
    journal = {Review of Scientific Instruments},
    volume = {81},
    number = {10},
    pages = {10D740},
    year = {2010},
    month = {10},
    issn = {0034-6748},
    doi = {10.1063/1.3496991},
    url = {https://doi.org/10.1063/1.3496991}
}

@article{Yun_2010,
    author = {Yun, G. S. and Lee, W. and Choi, M. J. and Kim, J. B. and Park, H. K. and Domier, C. W. and Tobias, B. and Liang, T. and Kong, X. and Luhmann, N. C., Jr. and Donné, A. J. H.},
    title = {Development of KSTAR ECE imaging system for measurement of temperature fluctuations and edge density fluctuations},
    journal = {Review of Scientific Instruments},
    volume = {81},
    number = {10},
    pages = {10D930},
    year = {2010},
    month = {10},
    issn = {0034-6748},
    doi = {10.1063/1.3483209},
    url = {https://doi.org/10.1063/1.3483209}
}

@article{Yang_2020,
doi = {10.1088/1741-4326/aba1cf},
url = {https://doi.org/10.1088/1741-4326/aba1cf},
year = {2020},
month = {aug},
publisher = {IOP Publishing},
volume = {60},
number = {9},
pages = {096023},
author = {Yang, S.M. and Park, J.-K. and Logan, N.C. and Zhu, C. and Hu, Q. and Jeon, Y.M. and In, Y. and Ko, W.H. and Kim, S.K. and Lee, Y.H. and Na, Y.S.},
title = {Localizing resonant magnetic perturbations for edge localized mode control in KSTAR},
journal = {Nuclear Fusion}
}

@article{Nam_2026,
doi = {10.1088/1741-4326/ae8283},
url = {https://doi.org/10.1088/1741-4326/ae8283},
year = {2026},
month = {jul},
publisher = {IOP Publishing},
volume = {66},
number = {11},
pages = {116015},
author = {Nam, Yong Un and Han, H. and Lee, H.H. and others},
title = {Overview of the KSTAR experiments and future plan},
journal = {Nuclear Fusion}
}

@article{In_2019,
doi = {10.1088/1741-4326/ab4631},
url = {https://doi.org/10.1088/1741-4326/ab4631},
year = {2019},
month = {oct},
publisher = {IOP Publishing},
volume = {59},
number = {12},
pages = {126045},
author = {In, Yongkyoon and Loarte, A. and Lee, H.H. and Kim, K. and Jeon, Y.M. and Park, J.-K. and Ahn, J.-W. and Park, G.Y. and Kim, M. and Park, H. and the 3D Physics Task Force in KSTAR},
title = {Test of the ITER-like resonant magnetic perturbation configurations for edge-localized mode crash suppression on KSTAR},
journal = {Nuclear Fusion}
}

@article{Park_2007,
    author = {Park, Jong-kyu and Boozer, Allen H. and Glasser, Alan H.},
    title = {Computation of three-dimensional tokamak and spherical torus equilibria},
    journal = {Physics of Plasmas},
    volume = {14},
    number = {5},
    pages = {052110},
    year = {2007},
    month = {05},
    issn = {1070-664X},
    doi = {10.1063/1.2732170}
}

@article{Hinton_1976,
  title = {Theory of plasma transport in toroidal confinement systems},
  author = {Hinton, F. L. and Hazeltine, R. D.},
  journal = {Rev. Mod. Phys.},
  volume = {48},
  issue = {2},
  pages = {239--308},
  numpages = {0},
  year = {1976},
  month = {Apr},
  publisher = {American Physical Society},
  doi = {10.1103/RevModPhys.48.239},
  url = {https://link.aps.org/doi/10.1103/RevModPhys.48.239}
}

@article{Rechester_1978,
  title = {Electron Heat Transport in a Tokamak with Destroyed Magnetic Surfaces},
  author = {Rechester, A. B. and Rosenbluth, M. N.},
  journal = {Phys. Rev. Lett.},
  volume = {40},
  issue = {1},
  pages = {38--41},
  numpages = {0},
  year = {1978},
  month = {Jan},
  publisher = {American Physical Society},
  doi = {10.1103/PhysRevLett.40.38},
  url = {https://link.aps.org/doi/10.1103/PhysRevLett.40.38}
}

@article{Chirikov_1979,
title = {A universal instability of many-dimensional oscillator systems},
journal = {Physics Reports},
volume = {52},
number = {5},
pages = {263-379},
year = {1979},
issn = {0370-1573},
doi = {https://doi.org/10.1016/0370-1573(79)90023-1},
url = {https://www.sciencedirect.com/science/article/pii/0370157379900231},
author = {Boris V Chirikov}
}

@article{Biewer_2003,
  title = {Electron Heat Transport Measured in a Stochastic Magnetic Field},
  author = {Biewer, T. M. and Forest, C. B. and Anderson, J. K. and Fiksel, G. and Hudson, B. and Prager, S. C. and Sarff, J. S. and Wright, J. C. and Brower, D. L. and Ding, W. X. and Terry, S. D.},
  journal = {Phys. Rev. Lett.},
  volume = {91},
  issue = {4},
  pages = {045004},
  numpages = {4},
  year = {2003},
  month = {Jul},
  publisher = {American Physical Society},
  doi = {10.1103/PhysRevLett.91.045004},
  url = {https://link.aps.org/doi/10.1103/PhysRevLett.91.045004}
}

@article{Guo_2022,
doi = {10.1088/1361-6587/ac93b0},
url = {https://doi.org/10.1088/1361-6587/ac93b0},
year = {2022},
month = {oct},
publisher = {IOP Publishing},
volume = {64},
number = {12},
pages = {124001},
author = {Guo, Weixin and Jiang, Min and Diamond, Patrick H and Chen, Chang-Chun and Cao, Mingyun and Li, Hanhui and Long, Ting},
title = {Theory of mean E \times B shear in a stochastic magnetic field: ambipolarity breaking and radial current},
journal = {Plasma Physics and Controlled Fusion}
}

@article{Mordijck_2016,
doi = {10.1088/0741-3335/58/1/014003},
url = {https://doi.org/10.1088/0741-3335/58/1/014003},
year = {2015},
month = {oct},
publisher = {IOP Publishing},
volume = {58},
number = {1},
pages = {014003},
author = {Mordijck, S and Rhodes, T L and Zeng, L and Doyle, E J and Schmitz, L and Chrystal, C and Strait, T J and Moyer, R A},
title = {Effects of resonant magnetic perturbations on turbulence and transport in DIII-D L-mode plasmas},
journal = {Plasma Physics and Controlled Fusion}
}

@article{Yoshida_2025,
doi = {10.1088/1741-4326/ad8ced},
url = {https://doi.org/10.1088/1741-4326/ad8ced},
year = {2025},
month = {feb},
publisher = {IOP Publishing},
volume = {65},
number = {3},
pages = {033001},
author = {Yoshida, M. and McDermott, R.M. and Angioni, C. and Camenen, Y. and Citrin, J. and Jakubowski, M. and Hughes, J.W. and Idomura, Y. and Mantica, P. and Mariani, A. and Mordijck, S. and Paul, E.J. and Tala, T. and Verdoolaege, G. and Zocco, A. and Casson, F.J. and Dif-Pradalier, G. and Duval, B. and Grierson, B.A. and Kaye, S.M. and Manas, P. and Maslov, M. and Odstrcil, T. and Rice, J.E. and Schmitz, L. and Sciortino, F. and Solano, E.R. and Staebler, G. and Valovič, M. and Wolfrum, E. and Snipes, J.A. and the Transport and Confinement Group},
title = {Transport and confinement physics: Chapter 2 of the special issue: on the path to tokamak burning plasma operation},
journal = {Nuclear Fusion}
}

@article{Conway_2015,
doi = {10.1088/0741-3335/57/1/014035},
url = {https://doi.org/10.1088/0741-3335/57/1/014035},
year = {2014},
month = {nov},
publisher = {IOP Publishing},
volume = {57},
number = {1},
pages = {014035},
author = {Conway, G D and Fietz, S and Müller, H W and Lunt, T and Simon, P and Suttrop, W and Maraschek, M and Happel, T and Viezzer, E and the ASDEX Upgrade Team},
title = {Impact of magnetic perturbation coils on the edge radial electric field and turbulence in ASDEX Upgrade},
journal = {Plasma Physics and Controlled Fusion}
}

\end{document}